\documentclass[proceedings]{stacs}
\stacsheading{2010}{585-596}{Nancy, France}
\firstpageno{585}

\usepackage[english]{babel}
\usepackage{german}
\usepackage{amssymb}
\usepackage{amsfonts}
\usepackage{amsmath}
\usepackage{graphicx}
\usepackage{subfigure}
\usepackage{algorithm}
\usepackage[noend]{algorithmic}
\usepackage{amsmath,amssymb,cite,wasysym}

\begin{document}

\title[The Recognition of Tolerance and Bounded Tolerance Graphs]{The Recognition of Tolerance \\
and Bounded Tolerance Graphs}

\author[lab1]{G.B. Mertzios}{George B. Mertzios}
\address[lab1]{Department of Computer Science, RWTH Aachen University, Aachen, Germany}  
\email{mertzios@cs.rwth-aachen.de} 

\author[lab2]{I. Sau}{Ignasi Sau}
\address[lab2]{Department of Computer Science, Technion, Haifa, Israel}	
\email{ignasi.sau@gmail.com}  

\author[lab3]{S. Zaks}{Shmuel Zaks}
\address[lab3]{Department of Computer Science, Technion, Haifa, Israel}	
\email{zaks@cs.technion.ac.il}


\begin{abstract}
Tolerance graphs model interval relations in such a way that intervals can
tolerate a certain degree of overlap without being in conflict. This subclass
of perfect graphs has been extensively studied, due to both its interesting
structure and its numerous applications. Several efficient algorithms for optimization
problems that are NP-hard on general graphs have been designed for tolerance
graphs. In spite of this, the recognition of tolerance graphs --~namely, the problem of 
deciding whether a given graph is a tolerance graph~-- 
as well as the recognition of their main subclass of bounded tolerance graphs, 
have been the most fundamental open problems on this class of graphs (cf.~the book on
tolerance graphs~\cite{GolTol04}) since their introduction in 1982~\cite{GoMo82}. 
In this article we prove that both recognition problems are NP-complete, 
even in the case where the input graph is a trapezoid graph. 
The presented results are surprising because, on the one hand, most subclasses of perfect graphs 
admit polynomial recognition algorithms 
and, on the other hand, bounded tolerance graphs were believed to be efficiently recognizable as 
they are a natural special case of trapezoid graphs (which can be recognized in polynomial time) 
and share a very similar structure with them. 
For our reduction we extend the notion of an \emph{acyclic orientation} of permutation and trapezoid graphs. 
Our main tool is a new algorithm that uses \emph{vertex splitting} to transform a given trapezoid graph 
into a permutation graph, while preserving this new acyclic orientation property. 
This method of vertex splitting is of independent interest; very recently, it has been proved a 
powerful tool also in the design of efficient recognition algorithms for other classes of graphs~\cite{MC-Trapezoid}.

\end{abstract}
\maketitle

\section{Introduction\label{sec:intro}}

\subsection{Tolerance graphs and related graph classes\label{classes}}

A simple undirected graph ${G=(V,E)}$ on $n$ vertices is a \emph{tolerance}
graph if there exists a collection ${I=\{I_{i}\ |\ i=1,2,\ldots ,n\}}$ of
closed intervals on the real line and a set ${t=\{t_{i}\ |\ i=1,2,\ldots ,n\}}$
of positive numbers, such that for any two vertices ${v_{i},v_{j}\in V}$, 
${v_{i}v_{j}\in E}$ if and only if ${|I_{i}\cap I_{j}|\geq \min \{t_{i},t_{j}\}}$. 
The pair ${\langle I,t\rangle}$ is called a \emph{tolerance representation}
of $G$. If $G$ has a tolerance representation ${\langle I,t\rangle}$, such
that ${t_{i}\leq |I_{i}|}$ for every $i=1,2,\ldots ,n$, then $G$ is called a 
\emph{bounded tolerance} graph and ${\langle I,t\rangle}$ a \emph{bounded
tolerance representation} of~$G$.

Tolerance graphs were introduced in~\cite{GoMo82}, in order to generalize
some of the well known applications of interval graphs. The main motivation
was in the context of resource allocation and scheduling problems, in which
resources, such as rooms and vehicles, can tolerate sharing among users~\cite%
{GolTol04}. If we replace in the definition of tolerance graphs the operator 
\emph{min} by the operator \emph{max}, we obtain the class of \emph{%
max-tolerance} graphs. Both tolerance and $\max $-tolerance graphs find in a
natural way applications in biology and bioinformatics, as in the comparison
of DNA sequences from different organisms or individuals~\cite{KKL+06}, by
making use of a software tool like BLAST~\cite{Altschul90}. Tolerance graphs
find numerous other applications in constrained-based temporal reasoning,
data transmission through networks to efficiently scheduling aircraft and
crews, as well as contributing to genetic analysis and studies of the 
brain~\cite{GolSi02,GolTol04}. This class of graphs has attracted many research 
efforts~\cite{GolTol04,GolumbicMonma84,GolSi02,HaSh04,Fel98,Bus06,KeBe04,BFI95,NaMa92,MSZ-Model-SIDMA-09}, 
as it generalizes in a natural way both interval graphs (when all
tolerances are equal) and permutation graphs (when $t_{i}=|I_{i}|$ for every 
$i=1,2,\ldots ,n$)~\cite{GoMo82}. For a detailed survey on tolerance graphs
we refer to~\cite{GolTol04}.

A \emph{comparability} graph is a graph which can be transitively oriented.
A \emph{co-comparability} graph is a graph whose complement is a
comparability graph. A \emph{trapezoid} (resp.~\emph{parallelogram} and 
\emph{permutation}) graph is the intersection graph of trapezoids
(resp.~parallelograms and line segments) between two parallel lines $L_{1}$
and $L_{2}$~\cite{Golumbic04}. Such a representation with trapezoids
(resp.~parallelograms and line segments) is called a \emph{trapezoid} (resp.~%
\emph{parallelogram} and \emph{permutation}) \emph{representation} of this
graph. A graph is bounded tolerance if and only if it is a parallelogram
graph~\cite{Langley93,BFI95}. Permutation graphs are a strict subset of parallelogram
graphs~\cite{classes99}. Furthermore, parallelogram graphs are a strict
subset of trapezoid graphs~\cite{Ryan98}, and both are subsets of
co-comparability graphs~\cite{GolTol04,Golumbic04}. On the contrary,
tolerance graphs are not even co-comparability graphs~\cite{GolTol04,Golumbic04}. 
Recently, we have presented in~\cite{MSZ-Model-SIDMA-09} a natural
intersection model for general tolerance graphs, given by parallelepipeds in
the three-dimensional space. This representation generalizes the
parallelogram representation of bounded tolerance graphs, and has been used
to improve the time complexity of minimum coloring, maximum clique, and
weighted independent set algorithms on tolerance graphs~\cite{MSZ-Model-SIDMA-09}.

Although tolerance and bounded tolerance graphs have been studied
extensively, the recognition problems for both these classes have been
the most fundamental open problems since their introduction in 1982~\cite{GolTol04,Golumbic04,BuIs07}.
Therefore, all existing algorithms assume that, along with the input tolerance graph, a tolerance representation of it
is given. The only result about the complexity of recognizing tolerance and
bounded tolerance graphs is that they have a (non-trivial) polynomial sized tolerance
representation, hence the problems of recognizing tolerance and bounded tolerance graphs
 are in the class NP~\cite{HaSh04}. Recently, a linear time recognition algorithm 
for the subclass of \emph{bipartite tolerance} graphs has been presented in~\cite{BuIs07}. 
Furthermore, the class of trapezoid graphs (which strictly contains parallelogram, 
i.e.~bounded tolerance, graphs~\cite{Ryan98}) 
can be also recognized in polynomial time~\cite{MaSpinrad94,Spinrad03,MC-Trapezoid}. 
On the other hand, the recognition of $\max$-tolerance graphs is known to be NP-hard~\cite{KKL+06}.
Unfortunately, the structure of $\max$-tolerance graphs differs significantly from that of 
tolerance graphs ($\max$-tolerance graphs are not even perfect, as they can contain induced~$C_5$'s~\cite{KKL+06}), 
so the technique used in~\cite{KKL+06} does not carry over to tolerance graphs.

Since very few subclasses of perfect graphs are known to be NP-hard to recognize, 
it was believed that the recognition of tolerance graphs was in P. 
Furthermore, as bounded tolerance graphs are equivalent to parallelogram graphs~\cite{Langley93,BFI95}, 
which constitute a natural subclass of trapezoid graphs and have a very similar structure, 
it was plausible that their recognition was also in P.

\subsection{Our contribution\label{contribution}}

In this article, we establish the complexity of 
recognizing tolerance and bounded tolerance graphs. 
Namely, we prove that both problems are surprisingly NP-complete, by providing a 
reduction from the monotone-Not-All-Equal-3-SAT (monotone-NAE-3-SAT) problem. 
Consider a boolean formula~$\phi $ in
conjunctive normal form with three literals in every clause (3-CNF), which
is monotone, i.e.~no variable is negated. The formula $\phi $ is called
NAE-satisfiable if there exists a truth assignment of the variables of $\phi 
$, such that every clause has at least one true variable and one false
variable. Given a monotone 3-CNF formula $\phi $, we construct a trapezoid
graph $H_{\phi }$, which is parallelogram, i.e.~bounded tolerance, if and
only if $\phi $ is NAE-satisfiable. Moreover, we prove that the constructed
graph $H_{\phi }$ is tolerance if and only if it is bounded tolerance. Thus,
since the recognition of tolerance and of bounded tolerance graphs are in
the class NP~\cite{HaSh04}, it follows that these problems are both
NP-complete. Actually, our results imply that the recognition problems
remain NP-complete even if the given graph is trapezoid, since the
constructed graph $H_{\phi }$ is trapezoid.

For our reduction we extend the notion of an \emph{acyclic orientation} of permutation and trapezoid graphs. 
Our main tool is a new algorithm that transforms a given trapezoid graph into a permutation graph 
by \emph{splitting} some specific vertices, while preserving this new acyclic orientation property. 
One of the main advantages of this algorithm is its robustness, in the sense that the constructed 
permutation graph does not depend on any particular trapezoid representation of the input graph~$G$. 
Moreover, besides its use in the present paper, this approach based on splitting vertices has been 
recently proved a powerful tool also in the design of efficient recognition algorithms for other 
classes of graphs~\cite{MC-Trapezoid}. 

\medskip

\textbf{Organization of the paper.} We first present in Section~\ref%
{trapezoid} several properties of permutation and trapezoid graphs, as well
as the algorithm Split-$U$, which constructs a permutation graph from a
trapezoid graph. In Section~\ref{bounded} we present the reduction of the
monotone-NAE-3-SAT problem to the recognition of bounded tolerance graphs. 
In Section~\ref{unbounded} we prove that this reduction can be extended to
the recognition of general tolerance graphs. 
Finally, we discuss the presented results and further research directions in
Section~\ref{conclusions}. 
Some proofs have been omitted due to space limitations; 
a full version can be found in~\cite{MertziosSauZaksAIB0905}.

\section{Trapezoid graphs and representations\label{trapezoid}}

In this section we first introduce (in Section~\ref{acyclic}) the notion of
an \emph{acyclic representation} of permutation and of trapezoid graphs.
This is followed (in Section~\ref{sec:structure}) by some structural
properties of trapezoid graphs, which will be used in the sequel for the
splitting algorithm Split-$U$. Given a trapezoid graph $G$ and a vertex
subset $U$ of $G$ with certain properties, this algorithm constructs a
permutation graph $G^{\#}(U)$ with $2|U|$ vertices, which is independent on
any particular trapezoid representation of the input graph $G$.

\medskip

\textbf{Notation.} We consider in this article simple undirected and
directed graphs with no loops or multiple edges. In an undirected graph $G$,
the edge between vertices $u$ and $v$ is denoted by $uv$, and in this case $u
$ and $v$ are said to be \emph{adjacent} in $G$. If the graph $G$ is
directed, we denote by $uv$ the arc from $u$ to $v$. Given a graph ${G=(V,E)}$
and a subset ${S \subseteq V}$, $G[S]$ denotes the induced subgraph of $G$ on
the vertices in $S$, and we use $E[S]$ to denote $E(G[S])$. 
Whenever we deal with a trapezoid (resp.~permutation and bounded tolerance,
i.e.~parallelogram) graph, we will consider w.l.o.g.~a trapezoid
(resp.~permutation and parallelogram) representation, in which all endpoints
of the trapezoids (resp.~line segments and parallelograms) are distinct~\cite%
{FishburnTrotter99,GolTol04,IsaakNT03}. Given a permutation graph $P$ along
with a permutation representation $R$, we may not distinguish in the
following between a vertex of $P$ and the corresponding line segment in $R$,
whenever it is clear from the context. Furthermore, with a slight abuse of
notation, we will refer to the line segments of a permutation representation just as \emph{lines}.

\subsection{Acyclic permutation and trapezoid representations\label{acyclic}}

Let $P=(V,E)$ be a permutation graph and $R$ be a permutation representation of $P$. 
For a vertex $u\in V$, denote by $\theta _{R}(u)$ the angle of the line of $u$ 
with $L_{2}$ in $R$. The class of permutation graphs is the intersection
of comparability and co-comparability graphs~\cite{Golumbic04}. Thus, given
a permutation representation $R$ of $P$, we can define two partial orders $%
(V,<_{R})$ and $(V,\ll _{R})$ on the vertices of~$P$~\cite{Golumbic04}.
Namely, for two vertices $u$ and $v$ of $G$, $u<_{R}v$ if and only if $uv\in
E$ and $\theta _{R}(u)<\theta _{R}(v)$, while $u\ll _{R}v$ if and only if 
$uv\notin E$ and $u$ lies to the left of $v$ in $R$. The partial order 
$(V,<_{R})$ implies a transitive orientation $\Phi _{R}$ of $P$, 
such that~$uv\in \Phi _{R}$ whenever $u<_{R}v$.

Let $G=(V,E)$ be a trapezoid graph, and $R$ be a trapezoid representation of 
$G$, where for any vertex $u\in V$, the trapezoid corresponding to $u$ in $R$
is denoted by $T_{u}$. Since trapezoid graphs are also co-comparability
graphs~\cite{Golumbic04}, we can similarly define the partial order $%
(V,\ll_{R})$ on the vertices of $G$, such that $u\ll _{R}v$ if and only if $%
uv\notin E$ and $T_{u}$ lies completely to the left of $T_{v}$ in $R$. In
this case, we may denote also $T_{u}\ll_{R}T_{v}$. % instead of $u\ll _{R}v$. 

In a given trapezoid representation $R$ of a trapezoid graph $G$, we denote
by~$l(T_{u})$ and~$r(T_{u})$ the left and the right line of $T_{u}$ in $R$,
respectively. Similarly to the case of permutation graphs, we use the
relation $\ll _{R}$ for the lines $l(T_{u})$ and $r(T_{u})$, e.g.~${l(T_{u})\ll _{R}r(T_{v})}$ 
means that the line $l(T_{u})$ lies to the left of
the line $r(T_{v})$ in $R$. Moreover, if the trapezoids of all vertices of a
subset $S\subseteq V$ lie completely to the left (resp.~right) of the
trapezoid $T_{u}$ in $R$, we write $R(S)\ll _{R}T_{u}$ (resp.~$T_{u}\ll
_{R}R(S)$). Note that there are several trapezoid representations of a
particular trapezoid graph $G$. Given one such representation $R$, we can
obtain another one $R^{\prime }$ by \emph{vertical axis flipping} of $R$,
i.e.~$R^{\prime }$ is the mirror image of $R$ along an imaginary line
perpendicular to $L_{1}$ and $L_{2}$. Moreover, we can obtain another
representation $R^{\prime \prime }$ of $G$ by \emph{horizontal axis flipping}
of $R$, i.e.~$R^{\prime \prime }$ is the mirror image of $R$ along an
imaginary line parallel to $L_{1}$ and $L_{2}$. We will extensively use 
these two operations throughout the article.

\begin{definition}
\label{acyclic-permutation}Let $P$ be a permutation graph with $2n$ vertices 
$\{u_{1}^{1},u_{1}^{2},u_{2}^{1},u_{2}^{2},\ldots ,u_{n}^{1},u_{n}^{2}\}$.
Let $R$ be a permutation representation and $\Phi _{R}$ be the corresponding
transitive orientation of $P$. The simple directed graph $F_{R}$ is obtained
by merging $u_{i}^{1}$ and $u_{i}^{2}$ into a single vertex~$u_{i}$, for
every $i=1,2,\ldots ,n$, where the arc directions of $F_{R}$ are implied by
the corresponding directions in $\Phi _{R}$. Then, 
\begin{enumerate}
\item $R$ is an \emph{acyclic permutation representation with respect to} $%
\{u_{i}^{1},u_{i}^{2}\}_{i=1}^{n}$\footnote{%
To simplify the presentation, we use throughout the paper $%
\{u_{i}^{1},u_{i}^{2}\}_{i=1}^{n}$ to denote the set of $n$ unordered pairs $%
\{u_{1}^{1},u_{1}^{2}\}, \{u_{2}^{1},u_{2}^{2}\}, \ldots,
\{u_{n}^{1},u_{n}^{2}\}$.
}, if $F_{R}$ has no directed cycle, 
\item $P$ is an \emph{acyclic permutation graph with respect to} $%
\{u_{i}^{1},u_{i}^{2}\}_{i=1}^{n}$, if $P$ has an acyclic representation $R$
with respect to $\{u_{i}^{1},u_{i}^{2}\}_{i=1}^{n}$.
\end{enumerate}
\end{definition}

\begin{definition}
\label{acyclic-trapezoid}Let $G$ be a trapezoid graph with $n$ vertices and $%
R$ be a trapezoid representation of $G$. Let $P$ be the permutation graph
with $2n$ vertices corresponding to the left and right lines of the
trapezoids in $R$, $R_{P}$ be the permutation representation of $P$ induced
by $R$, and $\{u_{i}^{1},u_{i}^{2}\}$ be the vertices of $P$ that correspond
to the same vertex $u_{i}$ of $G$, $i=1,2,\ldots ,n$. Then, %\vspace{-1mm}
\begin{enumerate}
\item $R$ is an \emph{acyclic trapezoid representation}, if $R_{P}$ is an
acyclic permutation representation with respect to $\{u_{i}^{1},u_{i}^{2}%
\}_{i=1}^{n}$, 
\item $G$ is an \emph{acyclic trapezoid graph}, if it has an acyclic
representation $R$.
\end{enumerate}
\end{definition}

The following lemma follows easily from Definitions~\ref{acyclic-permutation}
and~\ref{acyclic-trapezoid}. 

\begin{lemma}
\label{par}Any parallelogram graph is an acyclic trapezoid graph.
\end{lemma}

\subsection{Structural properties of trapezoid graphs}

\label{sec:structure}

In the following, we state some definitions concerning an arbitrary simple
undirected graph ${G=(V,E)}$, which are useful for our analysis. Although
these definitions apply to any graph, we will use them only for trapezoid
graphs. Similar definitions, for the restricted case where the graph $G$ is
connected, were studied in~\cite{Cheah96}. For $u\in V$ and $U\subseteq V$, $%
N(u)=\{v\in V\ |\ uv\in E\}$ is the set of adjacent vertices of $u$ in $G$, $%
{N[u]=N(u)\cup \{u\}}$, and ${N(U)=\bigcup_{u\in U}N(u)\setminus U}$. If ${%
N(U)\subseteq N(W)}$ for two vertex subsets $U$ and $W$, then $U$ is said to
be \emph{neighborhood dominated} by $W$. Clearly, the relationship of
neighborhood domination is transitive. 

Let ${C_{1},C_{2},\ldots ,C_{\omega },\ \omega \geq 1,}$ be the connected
components of ${G\setminus N[u]}$ and ${V_{i}=V(C_{i})}$, ${i=1,2,\ldots
,\omega }$. For simplicity of the presentation, we will identify in the
sequel the component $C_{i}$ and its vertex set $V_{i}$, ${i=1,2,\ldots
,\omega }$. For ${i=1,2,\ldots ,\omega }$, the \emph{neighborhood domination
closure} of $V_{i}$ with respect to $u$ is the set ${D_{u}(V_{i})=\{V_{p}\
|\ N(V_{p})\subseteq N(V_{i}),\ p=1,2,\ldots ,\omega \}}$ of connected
components of ${G\setminus N[u]}$. A component $V_{i}$ is called 
a \emph{master component} of~$u$ if ${|D_{u}(V_{i})|\geq |D_{u}(V_{j})|}$ for 
all ${j=1,2,\ldots ,\omega }$. The \emph{closure complement} of the neighborhood
domination closure ${D_{u}(V_{i})}$ is the set ${D_{u}^{\ast
}(V_{i})=\{V_{1},V_{2},\ldots ,V_{\omega }\}\setminus D_{u}(V_{i})}$.
Finally, for a subset ${S\subseteq \{V_{1},V_{2},\ldots ,V_{\omega }\}}$, a
component $V_{j} \in S$ is called \emph{maximal} if there is no component~$V_{k}\in S$ 
such that ${N(V_{j})\subsetneqq N(V_{k})}$.

\begin{figure}[tbh]
\centering
\includegraphics[width=0.82\textwidth]{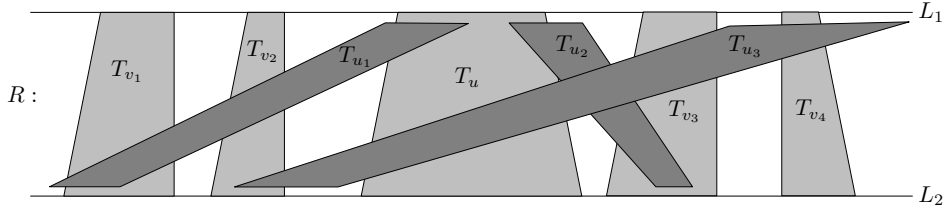}
\caption{A trapezoid representation $R$ of a trapezoid graph $G$.}
\label{counterexample}
\end{figure}

For example, consider the trapezoid graph $G$ with vertex set 
$\{u,u_{1},u_{2},u_{3},v_{1},v_{2},v_{3},v_{4}\}$, which is given by the
trapezoid representation $R$ of Figure~\ref{counterexample}. The connected
components of ${G\setminus N[u]=\{v_{1},v_{2},v_{3},v_{4}\}}$ are 
${V_{1}=\{v_{1}\}}$, ${V_{2}=\{v_{2}\}}$, ${V_{3}=\{v_{3}\}}$, and ${V_{4}=\{v_{4}\}}$. 
Then, ${N(V_{1})=\{u_{1}\}}$, ${N(V_{2})=\{u_{1},u_{3}\}}$, 
${N(V_{3})=\{u_{2},u_{3}\}}$, and ${N(V_{4})=\{u_{3}\}}$. Hence, 
${D_{u}(V_{1})=\{V_{1}\}}$, ${D_{u}(V_{2})=\{V_{1},V_{2},V_{4}\}}$, 
${D_{u}(V_{3})=\{V_{3},V_{4}\}}$, and ${D_{u}(V_{4})=\{V_{4}\}}$; thus, $V_{2}$
is the only master component of $u$. 
Furthermore, ${D_{u}^{\ast}(V_{1})=\{V_{2},V_{3},V_{4}\}}$, 
${D_{u}^{\ast }(V_{2})=\{V_{3}\}}$, ${D_{u}^{\ast}(V_{3})=\{V_{1},V_{2}\}}$, 
and ${D_{u}^{\ast}(V_{4})=\{V_{1},V_{2},V_{3}\}}$.

\begin{lemma}
\label{nonempty}Let $G$ be a simple graph, $u$ be a vertex of $G$, and let 
$V_{1},V_{2},\ldots ,V_{\omega },\ \omega \geq 1$, be the connected
components of ${G\setminus N[u]}$. If $V_{i}$ is a master component of $u$,
such that ${D_{u}^{\ast }(V_{i})\neq \emptyset }$, then ${D_{u}^{\ast
}(V_{j})\neq \emptyset}$ for every component $V_{j}$ of $G\setminus N[u]$.
\end{lemma}

\medskip

In the following we investigate several properties of trapezoid graphs, in
order to derive the vertex-splitting algorithm Split-$U$ in Section~\ref{splitting-algorithm}.

\begin{remark}
\label{mistake}Similar properties of trapezoid graphs have been studied in~%
\cite{Cheah96}, leading to another vertex-splitting algorithm, called
Split-All. However, the algorithm proposed in~\cite{Cheah96} is incorrect,
since it is based on an incorrect property\footnote{%
In Observation 3.1(5) of~\cite{Cheah96}, it is claimed that for an arbitrary
trapezoid representation $R$ of a connected trapezoid graph $G$, where $V_{i}
$ is a master component of $u$ such that ${D_{u}^{\ast}(V_{i})\neq
\emptyset }$ and ${R(V_{i})\ll _{R}T_{u}}$, it holds ${R(D_{u}(V_{i}))\ll
_{R}T_{u}\ll _{R}R(D_{u}^{\ast }(V_{i}))}$. However, the first part of the
latter inequality is not true. For instance, in the trapezoid graph $G$ of
Figure~\ref{counterexample}, ${V_{2}=\{v_{2}\}}$ is a master component of $u$, 
where $D_{u}^{\ast}(V_{2})=\{V_{3}\}=\{\{v_{3}\}\}{\neq \emptyset }$ and ${R(V_{2})\ll _{R}T_{u}}$. 
However, $V_{4}=\{v_{4}\}\in D_{u}(V_{2})$ and ${T_{u}\ll _{R}}T_{v_{4}}$,
and thus, ${R(D_{u}(V_{2}))\not\ll _{R}T}_{u} $.
}, as was also verified by~\cite{Communication09}. In the sequel of this section, we present new
definitions and properties. In the cases where a similarity arises with
those of~\cite{Cheah96}, we refer to it specifically.
\end{remark}

\begin{lemma}
\label{master}Let $R$ be a trapezoid representation of a trapezoid graph $G$, 
and $V_{i}$ be a master component of a vertex $u$ of $G$, such that 
$R(V_{i}){\ll }_{R}{T_{u}}$. Then, $T_{u}{\ll}_{R}R(V_{j})$ for every
component $V_{j}\in D_{u}^{\ast }(V_{i})$.
\end{lemma}

\begin{definition}
\label{Du}Let $G$ be a trapezoid graph, $u$ be a vertex of $G$, and $V_{i}$
be an arbitrarily chosen master component of $u$. Then, $\delta_{u}=V_{i}$ and
\begin{enumerate}
\item if $D_{u}^{\ast }(V_{i})=\emptyset $, then $\delta_{u}^{\ast}=\emptyset $.
\item if $D_{u}^{\ast }(V_{i})\neq \emptyset $, then $\delta_{u}^{\ast }={V_{j}}$, 
for an arbitrarily chosen maximal component $V_{j}\in {D_{u}^{\ast }(V_{i})}$.
\end{enumerate}
\end{definition}

Actually, as we will show in Lemma~\ref{separation-corollary}, 
the arbitrary choice of the components $V_{i}$ and $V_{j}$ in Definition~\ref{Du} 
does not affect essentially the structural properties of $G$ that we will 
investigate in the sequel.
From now on, whenever we speak about $\delta_{u}$ and $\delta_{u}^{\ast}$, 
we assume that these arbitrary choices of $V_{i}$ and $V_{j}$ have been already made.

\begin{definition}
\label{vertex-sets}Let $G$ be a trapezoid graph and $u$ be a vertex of $G$.
The vertices of $N(u)$ are partitioned into four possibly empty sets:
\begin{enumerate}
\item $N_{0}(u)$: vertices not adjacent to either ${\delta_{u}}$ or ${\delta_{u}^{\ast}}$.
\item $N_{1}(u)$: vertices adjacent to ${\delta_{u}}$ but not to ${\delta_{u}^{\ast }}$.
\item $N_{2}(u)$: vertices adjacent to ${\delta_{u}^{\ast }}$ but not to ${\delta_{u}}$.
\item $N_{12}(u)$: vertices adjacent to both ${\delta_{u}}$ and ${\delta_{u}^{\ast }}$.
\end{enumerate}
\end{definition}

In the following definition we partition the neighbors of a vertex of a
trapezoid graph~$G$ into four possibly empty sets. 
Note that these sets depend on a given trapezoid representation~$R$ of~$G$, in contrast
to the four sets of Definition~\ref{vertex-sets} that depend only on the graph~$G$ itself. 

\begin{definition}
\label{left-right}Let $G$ be a trapezoid graph, $R$ be a representation of $%
G $, and $u$ be a vertex of $G$. Denote by $D_{1}(u,R)$ and $D_{2}(u,R)$ the
sets of trapezoids of $R$ that lie completely to the left and to the right
of $T_{u}$ in $R$, respectively. Then, the vertices of $N(u)$ are
partitioned into four possibly empty sets:
\begin{enumerate}
\item $N_{0}(u,R)$: vertices not adjacent to either $D_{1}(u,R)$ or $D_{2}(u,R)$.
\item $N_{1}(u,R)$: vertices adjacent to $D_{1}(u,R)$ but not to $D_{2}(u,R)$.
\item $N_{2}(u,R)$: vertices adjacent to $D_{2}(u,R)$ but not to $D_{1}(u,R)$.
\item $N_{12}(u,R)$: vertices adjacent to both $D_{1}(u,R)$ and $D_{2}(u,R)$.
\end{enumerate}
\end{definition}

Suppose now that $\delta_{u}^{\ast} \neq \emptyset$, and let $V_{i}$ be the master component of $u$ that 
corresponds to $\delta_{u}$, cf.~Definition~\ref{Du}. 
Then, given any trapezoid representation $R$ of $G$, we may assume w.l.o.g.~that $R(V_{i}){\ll }_{R}{T_{u}}$, 
by possibly performing a vertical axis flipping of $R$.
The following lemma connects Definitions~\ref{vertex-sets} and~\ref{left-right}; 
in particular, it states that, if $R(V_{i}) \ll_{R} T_{u}$, then the partitions of the set $N(u)$ defined in 
these definitions coincide. 
This lemma will enable us to use in the vertex splitting (cf.~Definition~\ref{splitting}) 
the partition of the set $N(u)$ defined in Definition~\ref{vertex-sets}, 
independently of any trapezoid representation $R$ of $G$, and regardless of any particular 
connected components $V_{i}$ and $V_{j}$ of $G\setminus N[u]$.

\begin{lemma}
\label{separation-corollary}Let $G$ be a trapezoid graph, $R$ be a representation of $G$, 
and $u$ be a vertex of~$G$ with ${\delta_{u}^{\ast}\neq \emptyset}$. 
Let $V_{i}$ be the master component of $u$ that corresponds to $\delta_{u}$. 
If ${R(V_{i}){\ll }_{R}{T_{u}}}$, then ${N_{X}(u)=N_{X}(u,R)}$ for every $X\in \{0,1,2,12\}$.
\end{lemma}

\subsection{A splitting algorithm\label{splitting-algorithm}}

We define now the splitting of a vertex $u$ of a trapezoid graph $G$, where 
${\delta_{u}^{\ast }\neq \emptyset}$. 
Note that this splitting operation does not depend on any trapezoid representation of $G$.
Intuitively, if the graph~$G$ was given along with a specific trapezoid representation $R$, 
this would have meant that we replace the trapezoid $T_{u}$ in $R$ by its two lines $l(T_{u})$ 
and $r(T_{u})$.%\vspace{-1mm}

\begin{definition}
\label{splitting}Let $G$ be a trapezoid graph and $u$ be a vertex of $G$,
where ${\delta_{u}^{\ast}\neq \emptyset}$. The graph $G^{\#}(u)$ obtained by
the \emph{vertex splitting} of $u$ is defined as follows:
\begin{enumerate}
\item $V(G^{\#}(u))=V(G)\setminus \{u\}\cup \{u_{1},u_{2}\}$, where $u_{1}$
and $u_{2}$ are the two new vertices.
\item $E(G^{\#}(u))=E[V(G)\setminus \{u\}]\cup \{u_{1}x\ |\ x\in
N_{1}(u)\}\cup \{u_{2}x\ |\ x\in N_{2}(u)\}\cup \{u_{1}x,u_{2}x\ |\ x\in
N_{12}(u)\}$.
\end{enumerate}
The vertices $u_{1}$ and $u_{2}$ are the \emph{derivatives} of vertex $u$.
\end{definition}

We state now the notion of a standard trapezoid representation with respect
to a particular vertex. 

\begin{definition}
\label{standard}Let $G$ be a trapezoid graph and $u$ be a vertex of $G$,
where ${\delta_{u}^{\ast}\neq \emptyset}$. A trapezoid representation $R$ of 
$G$ is \emph{standard with respect to} $u$, if the following properties are
satisfied:
\begin{enumerate}
\item $l(T_{u})\ll_{R}R(N_{0}(u)\cup N_{2}(u))$.
\item $R(N_{0}(u)\cup N_{1}(u))\ll _{R}r(T_{u})$.
\end{enumerate}
\end{definition}

Now, given a trapezoid graph $G$ and a vertex subset $U=\{u_{1},u_{2},%
\ldots,u_{k}\}$, such that $\delta_{u_{i}}^{\ast }\neq \emptyset $ for every 
$i=1,2,\ldots ,k$, Algorithm Split-$U$ returns a graph $G^{\#}(U)$ by
splitting every vertex of~$U$ exactly once. At every step, Algorithm Split-$U
$ splits a vertex of $U$, and finally, it removes all vertices of the set $%
V(G)\setminus U$, which have not been split.

\begin{algorithm}[t]
\caption{Split-$U$} \label{alg1}
\begin{algorithmic} 
\REQUIRE{A trapezoid graph $G$ and a vertex subset $U=\{u_{1},u_{2},\ldots,u_{k}\}$, 
such that $\delta_{u_{i}}^{*}\neq\emptyset$ for all $i=1,2,\ldots, k$} 
\ENSURE{The permutation graph $G^{\#}(U)$\vspace{0.2cm}}

\STATE{$\overline{U} \leftarrow V(G) \setminus U$; $H_{0} \leftarrow G$\vspace{0.1cm}}

\FOR {$i=1$ to $k$}
     \STATE{$H_{i} \leftarrow H_{i-1}^{\#}(u_{i})$} 
\COMMENT{$H_{i}$ is obtained by the vertex splitting of $u_{i}$ in $H_{i-1}$\vspace{0.1cm}}
\ENDFOR 
\STATE{$G^{\#}(U) \leftarrow H_{k}[V(H_{k})\setminus \overline{U}]$\vspace{0.1cm}}
\COMMENT{remove from $H_{k}$ all unsplitted vertices}

\RETURN{$G^{\#}(U)$}
\end{algorithmic}
\end{algorithm}

\begin{remark}
\label{mistake-2}As mentioned in Remark~\ref{mistake}, a similar algorithm,
called Split-All, was presented in~\cite{Cheah96}. We would like to
emphasize here the following four differences between the two algorithms.
First, that Split-All gets as input a sibling-free graph $G$ (two vertices~$u,v$ 
of a graph $G$ are called \emph{siblings}, if $N[u]=N[v]$; $G$ is
called \emph{sibling-free} if $G$ has no pair of sibling vertices), while
our Algorithm Split-$U$ gets as an input any graph (though, we will use it
only for trapezoid graphs), which may contain also pairs of sibling vertices.
Second, Split-All splits all the vertices of the input graph, while Split-$U$
splits only a subset of them, which satisfy a special property. Third, the
order of vertices that are split by Split-All depends on a certain property
(inclusion-minimal neighbor set), while Split-$U$ splits the vertices in an
arbitrary order. Last, the main difference between these two algorithms is
that they perform a different vertex splitting operation at every step,
since Definitions~\ref{Du} and~\ref{vertex-sets} do not comply with the
corresponding Definitions 4.1 and 4.2 of~\cite{Cheah96}.
\end{remark}

\begin{theorem}
\label{split} Let $G$ be a trapezoid graph and $U=\{u_{1},u_{2},\ldots
,u_{k}\}$ be a vertex subset of $G$, such that $\delta_{u_{i}}^{\ast }\neq
\emptyset $ for every $i=1,2,\ldots ,k$. Then, the graph $G^{\#}(U)$
obtained by Algorithm Split-$U$, is a permutation graph with $2k$ vertices.
Furthermore, if $G$ is acyclic, then $G^{\#}(U)$ is acyclic with respect to $%
\{u_{i}^{1},u_{i}^{2}\}_{i=1}^{k}$, where $u_{i}^{1}$ and $u_{i}^{2}$ are
the derivatives of~$u_{i}$, $i=1,2,\ldots ,k$.
\end{theorem}

\section{The recognition of bounded tolerance graphs\label{bounded}}

In this section we provide a reduction from the 
\emph{monotone-Not-All-Equal-3-SAT (monotone-NAE-3-SAT)} problem 
to the problem of recognizing whether a given graph is a bounded tolerance graph. 
The problem of deciding whether a given monotone 3-CNF formula $\phi $ is 
NAE-satisfiable is known to be NP-complete. We can assume w.l.o.g.~that 
each clause has three distinct literals, i.e.~variables. Given a monotone 
3-CNF formula $\phi$, we construct in polynomial time a trapezoid graph $H_{\phi}$, 
such that $H_{\phi }$ is a bounded tolerance graph if and only if $\phi$ 
is NAE-satisfiable. To this end, we construct first a permutation graph 
$P_{\phi}$ and a trapezoid graph $G_{\phi}$.

\subsection{The permutation graph $P_{\protect\phi }$\label{P}}

\begin{figure}[htb]
\centering
\includegraphics[scale=0.7]{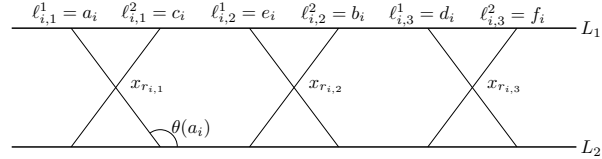}
\caption{The six lines of the permutation graph $P_{\protect\phi }$, which
correspond to the clause ${\protect\alpha _{i}=(x_{r_{i,1}}\vee
x_{r_{i,2}}\vee x_{r_{i,3}})}$ of the boolean formula $\protect\phi $.}
\label{clause}
\end{figure}

Consider a monotone 3-CNF formula ${\phi =\alpha _{1}\wedge \alpha
_{2}\wedge \ldots \wedge \alpha _{k}}$ with $k$ clauses and $n$ boolean
variables ${x_{1},x_{2},\ldots ,x_{n}}$, such that ${\alpha
_{i}=(x_{r_{i,1}}\vee x_{r_{i,2}}\vee x_{r_{i,3}})}$ for ${i=1,2,\ldots ,k}$, 
where ${1\leq r_{i,1}<r_{i,2}<r_{i,3}\leq n}$. We construct the
permutation graph $P_{\phi }$, along with a permutation representation $R_{P}$ 
of $P_{\phi }$, as follows. Let $L_{1}$ and $L_{2}$ be two parallel lines
and let $\theta (\ell )$ denote the angle of the line $\ell $ with $L_{2}$
in $R_{P}$. For every clause $\alpha _{i}$, ${i=1,2,\ldots ,k}$, we
correspond to each of the literals, i.e.~variables, $x_{r_{i,1}}$, 
$x_{r_{i,2}}$, and $x_{r_{i,3}}$ a pair of intersecting lines with 
endpoints on $L_{1}$ and $L_{2}$. 
Namely, we correspond to the variable $x_{r_{i,1}}$ the pair $\{{a_{i},c_{i}\}}$, 
to $x_{r_{i,2}}$ the pair $\{{e_{i},b_{i}\}}$ 
and to $x_{r_{i,3}}$ the pair $\{{d_{i},f_{i}\}}$, respectively, such that 
${\theta (a_{i})>\theta (c_{i})}$, 
${\theta (e_{i})>\theta (b_{i})}$, 
${\theta (d_{i})>\theta (f_{i})}$, and such that the lines 
${a_{i},c_{i}}$ lie completely to the left of ${e_{i},b_{i}}$ in $R_{P}$, and 
${e_{i},b_{i}}$ lie completely to the left of ${d_{i},f_{i}}$ in $R_{P}$, 
as it is illustrated in Figure~\ref{clause}. 
Denote the lines that correspond to the
variable $x_{r_{i,j}}$, $j=1,2,3$, by $\ell _{i,j}^{1} $ and $\ell _{i,j}^{2}
$, respectively, such that ${\theta (\ell _{i,j}^{1})>\theta (\ell
_{i,j}^{2})}$. That is, $(\ell _{i,1}^{1},\ell _{i,1}^{2})=(a_{i},c_{i})$, $%
(\ell _{i,2}^{1},\ell _{i,2}^{2})=(e_{i},b_{i})$, and $(\ell _{i,3}^{1},\ell
_{i,3}^{2})=(d_{i},f_{i})$. Note that no line of a pair $\{\ell
_{i,j}^{1},\ell _{i,j}^{2}\}$ intersects with a line of another pair $\{\ell
_{i^{\prime },j^{\prime }}^{1},\ell _{i^{\prime },j^{\prime }}^{2}\}$.

Denote by $S_{p}$, $p=1,2,\ldots ,n$, the set of pairs $\{\ell
_{i,j}^{1},\ell _{i,j}^{2}\}$ that correspond to the variable~$x_{p}$, i.e.~$%
r_{i,j}=p$. We order the pairs $\{\ell _{i,j}^{1},\ell _{i,j}^{2}\}$ such
that any pair of $S_{p_{1}}$ lies completely to the left of any pair of $%
S_{p_{2}}$, whenever $p_{1}<p_{2}$, while the pairs that belong to the same
set $S_{p}$ are ordered arbitrarily. For two consecutive pairs $\{\ell
_{i,j}^{1},\ell _{i,j}^{2}\}$ and $\{\ell_{i^{\prime },j^{\prime }}^{1},\ell
_{i^{\prime },j^{\prime }}^{2}\}$ in~$S_{p}$, where $\{\ell _{i,j}^{1},\ell
_{i,j}^{2}\}$ lies to the left of $\{\ell _{i^{\prime },j^{\prime
}}^{1},\ell _{i^{\prime },j^{\prime }}^{2}\}$, we add a pair $%
\{u_{i,j}^{i^{\prime },j^{\prime }},v_{i,j}^{i^{\prime },j^{\prime }}\}$ of
parallel lines that intersect both $\ell _{i,j}^{1}$ and $\ell _{i^{\prime
},j^{\prime }}^{1}$, but no other line. Note that $\theta (\ell
_{i,j}^{1})>\theta (u_{i,j}^{i^{\prime },j^{\prime }})$ and $\theta (\ell
_{i^{\prime },j^{\prime }}^{1})>\theta (u_{i,j}^{i^{\prime },j^{\prime }})$,
while $\theta (u_{i,j}^{i^{\prime },j^{\prime }})=\theta (v_{i,j}^{i^{\prime
},j^{\prime }})$. This completes the construction. Denote the resulting
permutation graph by $P_{\phi }$, and the corresponding permutation
representation of~$P_{\phi }$ by $R_{P}$. Observe that $P_{\phi }$ has $n$
connected components, which are called \emph{blocks}, one for each variable $%
x_{1},x_{2},\ldots ,x_{n}$.

\begin{figure}[tbh]
\centering
\includegraphics[width=0.8\textwidth]{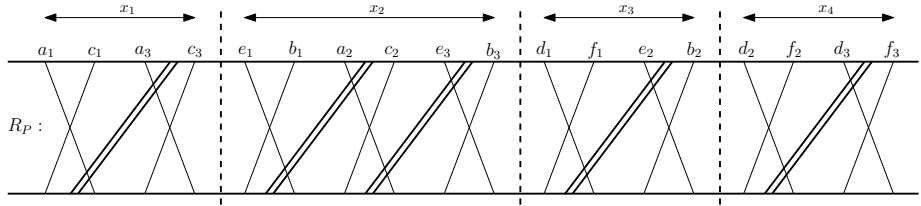}
\caption{The permutation representation $R_{P}$ of the permutation graph $P_{%
\protect\phi }$ for $\protect\phi =\protect\alpha _{1}\wedge \protect\alpha %
_{2}\wedge \protect\alpha_{3}= (x_{1}\vee x_{2}\vee x_{3})\wedge (x_{2}\vee
x_{3}\vee x_{4})\wedge (x_{1}\vee x_{2}\vee x_{4})$.}
\label{permutation}
\end{figure}

An example of the construction of $P_{\phi }$ and $R_{P}$ from $\phi $ with $%
k=3$ clauses and $n=4$ variables is illustrated in Figure~\ref{permutation}.
In this figure, the lines $u_{i,j}^{i^{\prime },j^{\prime }}$ and $v_{i,j}^{i^{\prime },j^{\prime }}$ are drawn in bold. 

The formula $\phi $ has $3k$ literals, and thus the permutation graph $%
P_{\phi }$ has $6k$ lines $\ell _{i,j}^{1},\ell _{i,j}^{2}$ in $R_{P}$, one
pair for each literal. Furthermore, two lines $u_{i,j}^{i^{\prime
},j^{\prime }},v_{i,j}^{i^{\prime },j^{\prime }}$ correspond to each pair of
consecutive pairs $\{\ell _{i,j}^{1},\ell _{i,j}^{2}\}$ and $\{\ell
_{i^{\prime },j^{\prime }}^{1},\ell _{i^{\prime },j^{\prime }}^{2}\}$ in $%
R_{P}$, except for the case where these pairs of lines belong to different
variables, i.e.~when $r_{i,j}\neq r_{i^{\prime },j^{\prime }}$. Therefore,
since $\phi $ has $n$ variables, there are $2(3k-n)=6k-2n$ lines $%
u_{i,j}^{i^{\prime },j^{\prime }},v_{i,j}^{i^{\prime },j^{\prime }}$ in $%
R_{P}$. Thus, $R_{P}$ has in total $12k-2n$ lines, i.e.~$P_{\phi }$ has $%
12k-2n$ vertices. In the example of Figure~\ref{permutation}, $k=3$, $n=4$,
and thus, $P_{\phi }$ has $28$ vertices.

Let $m=6k-n$, where $2m$ is the number of vertices in $P_{\phi }$. We group
the lines of $R_{P}$, i.e.~the vertices of $P_{\phi }$, into pairs $%
\{u_{i}^{1},u_{i}^{2}\}_{i=1}^{m}$, as follows. For every clause $\alpha _{i}
$, $i=1,2,\ldots ,k$, we group the lines $a_i,b_i,c_i,d_i,e_i,f_i$ into the
three pairs $\{a_{i},b_{i}\}$, $\{c_{i},d_{i}\}$, and $\{e_{i},f_{i}\}$. The
remaining lines are grouped naturally according to the construction; namely,
every two lines $\{u_{i,j}^{i^{\prime },j^{\prime }},v_{i,j}^{i^{\prime
},j^{\prime }}\}$ constitute a pair. 

\begin{lemma}
\label{2b-direction}If the permutation graph $P_{\phi }$ is acyclic with
respect to $\{u_{i}^{1},u_{i}^{2}\}_{i=1}^{m}$ then the formula $\phi $ is NAE-satisfiable.
\end{lemma}

The truth assignment $(x_{1},x_{2},x_{3},x_{4})=(1,1,0,0)$ is NAE-satisfying for 
the formula $\phi$ of Figure~\ref{permutation}. 
The acyclic permutation representation $R_{0}$ of~$P_{\phi}$ with respect 
to $\{u_{i}^{1},u_{i}^{2}\}_{i=1}^{m}$, which corresponds to this assignment, 
can be obtained from $R_{P}$ by performing a horizontal axis flipping 
of the two blocks that correspond to the variables $x_3$ and $x_4$, respectively.

\subsection{The trapezoid graphs $G_{\protect\phi }$ and $H_{\protect\phi }$%
\label{G-H}}

Let $\{u_{i}^{1},u_{i}^{2}\}_{i=1}^{m}$ be the pairs of vertices in the
permutation graph $P_{\phi }$ and $R_{P}$ be its permutation
representation. We construct now from $P_{\phi }$ the trapezoid graph $%
G_{\phi }$ with $m$ vertices $\{u_{1},u_{2},\ldots,u_{m}\}$, as follows. We
replace in the permutation representation $R_{P}$ for every $i=1,2,\ldots ,m$
the lines $u_{i}^{1}$ and $u_{i}^{2}$ by the trapezoid $T_{u_{i}}$, which
has $u_{i}^{1}$ and $u_{i}^{2}$ as its left and right lines, respectively.
Let $R_{G}$ be the resulting trapezoid representation of~$G_{\phi }$.

\begin{figure}[t]
\centering\includegraphics[scale=0.9]{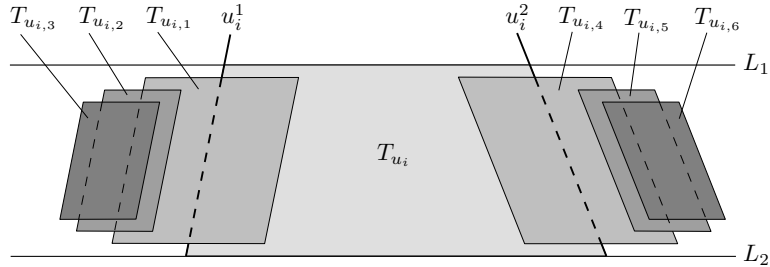} 
\caption{The addition of the six parallelograms $T_{u_{i,1}}, T_{u_{i,2}},
\ldots, T_{u_{i,6}}$ to the trapezoid $T_{u_{i}}$, ${i=1,2,\ldots ,m}$, in the
construction of the trapezoid graph $H_{\protect\phi}$ from $G_{\protect\phi}$.}
\label{extra-6}
\end{figure}

Finally, we construct from $G_{\phi }$ the trapezoid graph $H_{\phi}$ with $%
7m$ vertices, by adding to every trapezoid $T_{u_{i}}$, $i=1,2,\ldots ,m$,
six parallelograms $T_{u_{i,1}}, T_{u_{i,2}}, \ldots, T_{u_{i,6}}$ in the
trapezoid representation $R_{G}$, as follows. Let $\varepsilon$ be the
smallest distance in $R_{G}$ between two different endpoints on $L_{1}$, or
on $L_{2}$. The right (resp.~left) line of $T_{u_{1,1}}$ lies to the right
(resp.~left) of $u_{1}^{1}$, and it is parallel to it at distance $\frac{%
\varepsilon}{2}$. The right (resp.~left) line of $T_{u_{1,2}}$ lies to the
left of $u_{1}^{1}$, and it is parallel to it at distance $\frac{\varepsilon%
}{4}$ (resp.~$\frac{3\varepsilon}{4}$). Moreover, the right (resp.~left)
line of $T_{u_{1,3}}$ lies to the left of $u_{1}^{1}$, and it is parallel to
it at distance $\frac{3\varepsilon}{8}$ (resp.~$\frac{7\varepsilon}{8}$).
Similarly, the left (resp.~right) line of $T_{u_{1,4}}$ lies to the left
(resp.~right) of $u_{1}^{2}$, and it is parallel to it at distance $\frac{%
\varepsilon }{2}$. The left (resp.~right) line of $T_{u_{1,5}}$ lies to the
right of $u_{1}^{2}$, and it is parallel to it at distance $\frac{\varepsilon%
}{4}$ (resp.~$\frac{3\varepsilon}{4}$). Finally, the right (resp.~left) line
of $T_{u_{1,6}}$ lies to the right of $u_{1}^{2}$, and it is parallel to it
at distance $\frac{3\varepsilon}{8}$ (resp.~$\frac{7\varepsilon}{8}$), as
illustrated in Figure~\ref{extra-6}.

After adding the parallelograms $T_{u_{1,1}}, T_{u_{1,2}}, \ldots,
T_{u_{1,6}}$ to a trapezoid $T_{u_{1}}$, we update the smallest distance $%
\varepsilon $ between two different endpoints on $L_{1}$, or on $L_{2}$ in
the resulting representation, and we continue the construction iteratively
for all $i=2,\ldots ,m$. Denote by~$H_{\phi }$ the resulting trapezoid graph
with $7m$ vertices, and by $R_{H}$ the corresponding trapezoid representation. 
Note that in $R_{H}$, between the endpoints of the parallelograms 
$T_{u_{i,1}}$, $T_{u_{i,2}}$, and~$T_{u_{i,3}}$ 
(resp.~$T_{u_{i,4}}$, $T_{u_{i,5}}$, and $T_{u_{i,6}}$) on $L_{1}$ and $L_{2}$,
there are no other endpoints of $H_{\phi }$, except those of $u_{i}^{1}$ (resp.~$u_{i}^{2}$), 
for every $i=1,2,\ldots,m$. 
Furthermore, note that $R_{H}$ is standard with respect to $u_{i}$, for every $i=1,2,\ldots,m$.

\begin{theorem}
\label{bounded-both-directions}The formula $\phi$ is NAE-satisfiable if and
only if the trapezoid graph $H_{\phi}$ is a bounded tolerance graph.
\end{theorem}

For the sufficiency part of the proof of Theorem~\ref{bounded-both-directions},
the algorithm Split-All plays a crucial role. Namely, given the parallelogram graph $H_{\phi}$ 
(which is acyclic trapezoid by Lemma~\ref{par}), 
we construct with this algorithm the acyclic permutation graph $P_{\phi}$ 
and then a NAE-satisfying assignment of the formula $\phi$. 
Since monotone-NAE-3-SAT is NP-complete, the problem of recognizing 
bounded tolerance graphs is NP-hard by Theorem~\ref{bounded-both-directions}. 
Moreover, since this problem lies in NP~\cite{HaSh04}, 
we summarize our results as follows.
\begin{theorem}
\label{bounded-tolerance}Given a graph $G$, it is NP-complete to decide
whether it is a bounded tolerance graph.
\end{theorem}

\section{The recognition of tolerance graphs\label{unbounded}}

In this section we show that the reduction from the monotone-NAE-3-SAT
problem to the problem of recognizing bounded tolerance graphs presented in
Section~\ref{bounded}, can be extended to the problem of recognizing general
tolerance graphs. In particular, we prove that the constructed trapezoid graph 
$H_{\phi}$ is a tolerance graph if and only if it is a bounded tolerance graph. 
Then, the main result of this section follows.

\begin{theorem}
\label{tolerance}Given a graph $G$, it is NP-complete to decide whether it
is a tolerance graph. The problem remains NP-complete even if the given
graph $G$ is known to be a trapezoid graph.
\end{theorem}

\section{Concluding remarks\label{conclusions}}
In this article we proved that both tolerance and bounded tolerance graph
recognition problems are NP-complete, by providing a reduction from the
monotone-NAE-3-SAT problem, thus answering a longstanding open question.
The recognition of unit and of proper tolerance graphs, as well as of 
any other subclass of tolerance graphs, except bounded tolerance and 
bipartite tolerance graphs~\cite{BuIs07}, remain interesting open 
problems~\cite{GolTol04}.

{\footnotesize 
\bibliographystyle{abbrv}
\bibliography{Mertzios_Tolerance} 
}
\vspace{-1cm}

\end{document}